\documentclass[twocolumn,showpacs,preprintnumbers,amsmath,amssymb,floatfix]{revtex4}
\usepackage{graphicx}
\usepackage{dcolumn}
\usepackage{bm}
\usepackage{amssymb}
\usepackage{amsmath}

\begin{document}
\title{Noise-Driven Mechanism for Pattern Formation}
\author{J. Buceta$^{\dag}$ , M. Iba\~{n}es$^{\ddag}$ , J.
M. Sancho$^{\ddag}$, and Katja Lindenberg$^{\dag}$ }
\affiliation{\dag University of California San Diego, and Institute
for Nonlinear Science, La Jolla, CA 92093-0340, USA\\
\ddag Departament d'Estructura i Constituents de la Mat\`{e}ria,
Universitat de Barcelona, Diagonal 647, E-08028 Barcelona, Spain}

\begin{abstract}
We extend the mechanism for noise-induced phase transitions
proposed by Iba\~{n}es {\em et al.}
[{\em Phys. Rev. Lett.} {\bf 87},
020601-1 (2001)] to pattern formation phenomena.  In contrast with known
mechanisms for pure noise-induced pattern formation, this
mechanism is {\em not} driven by a short-time
instability amplified by collective effects. The phenomenon is analyzed by
means of a modulated mean field approximation and numerical simulations.
\end{abstract}

\maketitle

\section{Introduction}
\label{introduction}

Starting with the seminal work of Horsthemke and Mansour on the
Verhulst model~\cite {horsthemke1}, noise-induced phenomena
have been a subject of intense interest~\cite{horsthemke2}.  Much of the
early work dealt with noise-induced phenomena in
zero-dimensional systems.  More recently, it has become widely
recognized that the effects of fluctuations on systems
with a large number of degrees of freedom, so-called
{\em spatially extended} systems consisting of coupled
zero-dimensional systems, can be even more striking~\cite{ojalvo}.
First-order~\cite{muller,oliver}
and second-order~\cite{vandenbroeck1,vandenbroeck2,jordi-parrondo}
noise-induced phase transitions, noise-induced
patterns~\cite{jordiPRL,kramers,parrondo,zaikin0}, and doubly stochastic
resonance~\cite{zaikin} are examples that illustrate the broad interest
in the subject.  The noise intensity is the parameter that controls the
creation of order in all these processes.  The underlying
mechanism that drives the order in these extended systems
is a noise-induced short-time
instability that is amplified by collective effects that do not occur in
the absence of coupling or in the absence of noise.
Moreover, as the intensity of the fluctuations increases, a
re-entrance phenomenon occurs: the noise that first drove
the system to an ordered state now restores the disorder.
It is important to recall that this particular mechanism is in a sense
opposite to that responsible for noise-induced phenomena
in zero-dimensional systems~\cite{vandenbroeck1,vandenbroeck2}
in that these noise-induced phase transitions
are observed only when the associated zero-dimensional units exhibit no
interesting noise-induced behavior.  It was therefore thought for a long
time that coupling of zero-dimensional units that undergo a noise-induced
transition would exhibit no interesting collective effects.

However, Iba\~nes {\em et al.} \cite{ibanes} introduced a class of exactly
solvable models that exhibit {\em both} noise-induced transitions
in the zero-dimensional case {\em and}
noise-induced phase transitions in the associated extended system.
They stress that this phase transition
arises from an effective equilibrium potential in the steady state
and does not require a short-time instability or any other
reference to the short time behavior of the system. The key
ingredient is the combination of relaxations that rely on
field dependent kinetic coefficients and the disordering effects of external
fluctuations. Furthermore, in this case no re-entrance phenomenon occurs:
the system becomes more ordered with increasing noise intensity.

In this paper we extend this mechanism to
pattern formation. We modify the coupling term of the model in such a way
that a morphological instability can appear. We will show, by means of
numerical simulations and analytic calculations, that an
increasingly ordered spatial structure develops as the intensity of the
fluctuation increases beyond a critical value.  Moreover, we show that for
sufficiently high noise intensity there is non-monotonic behavior as a
function of the coupling, that is, there exists and optimal value of the
coupling for which the ordered structure exhibits maximum coherence.

The paper is organized as follows. We introduce the model in
Sec.~\ref{modelsec} and explicitly show that no short-time instability
can drive pattern formation in this system.  In
Sec.~\ref{modulated} and Appendix~\ref{a} we present a {\em modulated}
mean field theory and the phase diagram of the model obtained from this
theory.  Numerical simulations that confirm the
qualitative validity of the
theoretical results are presented in Sec.~\ref{numerical}. Finally, we
summarize our main conclusions in Sec.~\ref{conclusions}

\section{The model}
\label{modelsec}

As introduced in~\cite{ibanes}, we consider the following Langevin
dynamics for a space and time dependent
scalar field $\phi_{\bm r}\left(t\right)$:
\begin{equation}
\dot{\phi_{\bm r}}=\Gamma (\phi_{\bm r})
\left[-a\phi_{\bm r} +
{\mathcal L}\phi_{\bm r} \right] +\left[ \Gamma (\phi_{\bm r})\right]^{1/2}
\xi ({\bm r},t) .
\label{model}
\end{equation}
Here the local force $-a\phi$ arises from a monostable local
potential $a\phi^2/2$ and ${\mathcal L}$ stands for the spatial
coupling operator.  The
field-dependent kinetic coefficient is
\begin{equation}
\Gamma (\phi) =\frac{1}{ 1+c \phi^2}.
\end{equation}
The coupling of this kinetic coefficient with the noise favors
fluctuations in the disordered state $\phi=0$.
Both $a$ and $c$ are positive constants. The
noise term is assumed to be Gaussian with zero mean value
and correlation function
\begin{equation}
\langle \xi ( {\bm r},t) \xi ({\bm r}^{\prime},t^{\prime })\rangle
=2\sigma ^{2}\delta( t-t^{\prime}) \delta({\bm r}-{\bm r}^{\prime }),
\end{equation}
where the brackets $\langle~\rangle$ denote a statistical average.
The associated stochastic integral is interpreted in the Stratonovich
sense~\cite{horsthemke2}.

In the absence of coupling (${\mathcal L} =0$), the
model~(\ref{model}) reduces to a collection of zero-dimensional systems
that undergo a noise-induced transition \cite{horsthemke2}.  At small noise
intensity, the
stationary probability density has a single maximum at $\phi=0$.
As the noise intensity crosses the critical value $\sigma_c^2=a/c$ the
stationary probability density shows two symmetric maxima about a minimum at
$\phi=0$.  This transition involves no symmetry breaking since in
all cases the average $\langle\phi_{\bm r}\rangle=0$.
When sites are coupled diffusively, i.e., ${\mathcal L}
=D\nabla ^{2}$, the model~(\ref{model}) exhibits a noise-induced {\em phase}
transition \cite{ibanes}.  For a given value of the coupling, as
the intensity of the
fluctuations increases, the systems undergoes a second-order phase
transition that involves spontaneous symmetry breaking, that is, a
transition from a state with $\langle\phi_{\bm r}\rangle=0$ to one with
$\langle\phi_{\bm r}\rangle\neq 0$.  It is worth noting that
when the coupling strength parameter $D$ goes to infinity,
the location of that transition point is the
same as in a zero-dimensional ensemble,
$\sigma _{c}^{2}=a/c$ \cite{ibanes},
and that the symmetry-breaking in the coupled system can in that
case be understood in
terms of the dynamics associated with the effective potential in the
zero-dimensional case.

We show herein that more complex structural changes can be obtained
by modifying the coupling term ${\mathcal L}$. In particular,
pattern formation phenomena occur when the system develops a
morphological instability, that is, when a Fourier mode of wavevector
${\bm k}$ {\em other than} ${\bm k}={\bm 0}$ becomes
unstable~\cite{cross}.  Drawing parallels with
previous literature on noise-induced
phenomena~\cite{jordiPRL,kramers,parrondo,zaikin0}, we consider
a Swift-Hohenberg coupling, that is~\cite{swift},
\begin{equation}
{\mathcal L}=-D\left( k_{0}^{2}+\nabla
^{2}\right) ^{2}.
\end{equation}
The effect of this coupling can be deduced
by applying ${\mathcal L}$ to a plane wave $e^{i{\bm k} \cdot{\bm r}}$,
\begin{equation}
{\mathcal L}e^{i{\bm k}\cdot{\bm r}}=\omega ( k)
 e^{i{\bm k}\cdot{\bm r}},
\end{equation}
where $\omega (k) =-D(k_0^2-k^2)^2$ is the
(continuous) dispersion relation (we use bold for vectorial quantities and
italic for their magnitudes). Thus, the largest eigenvalues, $\omega(k)=0$,
are those associated with Fourier modes
of wavenumber magnitude $ k=k_0$.  We will see below that this behavior
leads to a morphological instability in the presence of noise.  We stress that
a morphological instability
is a key ingredient of the pattern formation mechanism, but the
specific functional form of the coupling term is not.

In the next two sections we show in detail that our model leads to
a noise-induced phase transition to patterned states.  We end this
section
by showing explicitly that the transition is {\em not}
caused by short-time
instabilities~\cite{jordiPRL,kramers,parrondo,zaikin0}.
The relevant short-time evolution equation
for the model can be obtained by averaging Eq.~(\ref{model})
and expanding for small $\phi_{\bm r}$ around the average
value $\langle \phi_{\bm r} \rangle=0$:
\begin{equation}
\dot{\phi}_{\bm r} =(-a-\sigma ^{2}c+{\mathcal L}) \phi_{\bm r} .
\end{equation}
Writing $\phi_{\bm r}(t)$ as a sum (in an infinite system as an integral)
over Fourier modes,
\begin{equation}
\phi_{\bm r}(t) =\sum_{\bm k}\tilde{\phi}_{\bm k}( t)
e^{i{\bm k}\cdot{\bm r}},
\end{equation}
and integrating, we obtain
\begin{equation}
\phi_{\bm r}(t)=\sum_{\bm k}\tilde{\phi}_{\bm k}
\exp\left\{\left[ -a-\sigma ^2c+\omega ( k) \right] t
+i{\bm k}\cdot{\bm r}\right\}.
\end{equation}
Since $\omega (k) \leq 0$, all modes decay and there is no short-time
instability.

\section{Modulated mean field theory}
\label{modulated}

To find the phase diagram for our noisy spatially extended system,
we introduce a {\em modulated} mean field theory.
In order to implement this theory and, in
particular, to calculate the effect of the coupling operator ${\mathcal L}$
in the mean field context, we need to explicitly
distinguish different locations ${\bm r}$ and ${\bm r}'$ in a way
that requires discretization of the system.  Since numerical
simulations also
involve discretization, this procedure does not interfere with the
comparisons of theoretical and numerical results.  With the understanding
of the action of the translation operator
\begin{equation}
\exp \left(\delta x\frac{\partial }{\partial x}\right)f(x) =f(x+\delta x),
\end{equation}
it is straightforward to deduce the discrete version of the
Swift-Hohenberg coupling operator,
\begin{equation}
\mathcal{L}=-D\left[ k_{0}^{2}+\left(\frac{2}{\Delta x}\right)^2
\sum_{i=1}^{d}
\sinh^2\left( \frac{\Delta x}{2}\frac{\partial }{\partial x_{i}}\right)
\right] ^{2},
\label{l-discrete}
\end{equation}
where $d$ stands for the spatial dimension, $\Delta x$ for the lattice
spacing, and $\frac{\partial }{\partial x_i}$ indicates a partial
derivative with respect the $i$-component of the position vector
$\bm r =(x_1,x_2,\ldots ,x_i,\ldots ,x_d ) $.
As in the continuous case, the discrete
dispersion relation can be obtained by applying
the operator (\ref{l-discrete}) to a plane wave
$e^{i{\bm k}\cdot{\bm r}}$, to obtain
\begin{equation}
\omega({\bm k}) =-D\left[k_0^2-\left(\frac{2}{\Delta x}\right)^2
\sum_{i=1}^d \sin^2 \left(\frac{\Delta x}{2}k_i \right)\right]^2.
\label{dispersion}
\end{equation}
Here $k_i$ denotes component $i$ of the wave vector ${\bm k}
=(k_1,k_2,\ldots,k_i,\ldots,k_d)$.

Note that as in the continuous problem, $\omega({\bm k})$ is nonpositive
for any value of $k$, but that in the discrete case it depends
not only on the magnitude but also on the direction of ${\bm k}$.
Of particular importance in our subsequent analysis are those modes
for which $\omega({\bm k})=0$.  In the continuum these are the modes with
$k=k_0$, which are all those that lie on a continuous hypersurface
in reciprocal space of radius
$k_0$ around the origin.  In the discretized system the magnitudes
$k^*$ of the most unstable modes are shifted from $k_0$ and depend on
direction, as can be seen by solving Eq.~(\ref{dispersion}).
The longest
vectors such that $\omega({\bm k}^*)=0$ lie along the Cartesian directions,
e.g. $(k^*,0,0,\ldots, 0)$ and have magnitude
\begin{equation}
\max k^{\ast}=\frac{2}{\Delta x}\arcsin\left(\frac{k_0\Delta x}{2}
\right).
\label{mink}
\end{equation}
The shortest lie along a reciprocal space diagonal, e.g.
$\frac{1}{\sqrt{d}}(k^*,k^*,k^*,\ldots,k^*)$, and have magnitude
\begin{equation}
\min k^{\ast}=\frac{2\sqrt{d}}{\Delta x}\arcsin\left(
\frac{k_0\Delta x}{2\sqrt{d}}\right).
\label{maxk}
\end{equation}
If $k_0\Delta x \leq 1$ (which it always will be in our analysis) then the
difference between these two values is smaller than 3\%.  It is therefore
only a mild approximation to neglect the directional
dependence of the solutions of $\omega({\bm k}^*)=0$ and focus on the
magnitude, $\omega(k^*)=0$.

To establish the existence of patterns of a characteristic length
scale, we seek a spatially periodic structure defined by wavevectors
${\bm k}$ whose magnitude $k$ is associated with the inverse of this
length scale.  As we shall see, the appropriate wavevectors to focus on are
precisely those of magnitude $k^*$, that is, those for which
$\omega({\bm k})=0$.  As is customary in mean field theories, we make
an ansatz about the behavior of the field at sites ${\bm r}'$ {\em other}
than the focus
point ${\bm r}$ which are coupled to it by the operator ${\mathcal L}$, one
that incorporates an appropriate spatial modulation:
\begin{equation}
\phi_{\bm r^\prime}={\mathcal A}(k^*) \sum_{\left\{\bm{k}^{\ast}\right\} }
\cos \left[
{\bm k}\cdot ({\bm r}-{\bm r}^\prime) \right]
\label{mean field}
\end{equation}
where the sum (or, in an infinite system, the integral) is
over wavevectors of magnitude $k^*$.  Our ansatz
thus also incorporates the assumption that all modes of this
magnitude contribute
with equal (direction-independent) weight ${\mathcal A}(k^*)$.
In the appendix we detail some of the steps that show the action of the
coupling operator on this ansatz, to arrive at
the result
\begin{equation}
\mathcal{L}\phi_{\bm{r}}=D_1\left[\mathfrak{n}(k^{\ast})
\mathcal{A}(k^{\ast})-\phi_{\bm{r}}\right]
\label{finalapprox}
\end{equation}
where
\begin{equation}
\mathfrak{n}(k^{\ast}) =\frac{d\pi^{d/2}}{\Gamma(d/2+1)}
\left(\frac{Nk^*}{2\pi}\right)^{d-1}
\label{frakn}
\end{equation}
is the number of modes of magnitude $k^*$, and
\begin{equation}
D_1=D\left[\left(\frac{2d}{(\Delta x)^2}
-k_0^2\right)^2+\frac{2d}{(\Delta x)^4}\right].
\end{equation}
Substitution in Eq.~(\ref{model}) then leads to an equation that depends
only on a generic site index ${\bm r}$ that can simply be dropped:
\begin{equation}
\begin{aligned}
\dot{\phi}=\Gamma (\phi)& \left\{-a\phi
+D_1(k^*) \left[ \mathfrak{n}(k^*){\cal A}(k^*) -\phi \right] \right\}
\\
& +\left[ \Gamma (\phi)\right]^{1/2}\xi(t),
\label{ec1}
\end{aligned}
\end{equation}
and $\xi(t)$ is zero-centered Gaussian noise $\delta$-correlated in time,
\begin{equation}
\langle \xi (t) \xi (t^{\prime })\rangle
=\frac{ 2\sigma^2}{(\Delta x)^d}\delta( t-t^{\prime}).
\end{equation}
Here we have incorporated the fact that the continuum delta function
$\delta({\bm r}-{\bm r}')$ has been replaced in the usual way by a ratio
that contains the Kronecker delta and the lattice spacing,
$\delta_{{\bm r},{\bm r}'}/(\Delta x)^2$.   Henceforth we set $\Delta x
=1$.

The amplitude ${\mathcal A}(k^*)$ is the mean field quantity that must be
chosen self-consistently to complete the solution of the problem.
The stationary probability density for the stochastic
process~(\ref{finalapprox}) is
\begin{equation}
\begin{aligned}
&\rho \left( \phi ; {\mathcal A}(k^*)\right)={\mathcal N}
\left[\mathfrak{n}(k^*){\mathcal A}(k^*)\right](1+c\phi^2)^{1/2}
\\
~~~~~~&\times \,\exp\left\{-\frac{1}{\sigma^2}
\left[\frac{1}{2}(a+D_1)\phi^2-D_1\mathfrak{n}(k^*)
{\mathcal A}(k^*)\phi \right]\right\},
\end{aligned}
\end{equation}
where the normalization constant
${\mathcal N}\left[\mathfrak{n}(k^*){\mathcal A}(k^*)\right]$
depends on the amplitude and must therefore be carefully included in the
self-consistent solution. Self-consistency
is then embodied in the assumption that $\mathfrak{n}(k^*){\mathcal A}(k^*)$
is the average value of the field at any point in space, i.e. in the
requirement that
\begin{equation}
\mathfrak{n}(k^*){\mathcal A}(k^*) =\int_{-\infty}^\infty
\phi \rho \left( \phi; {\mathcal A} (k^*) \right) d\phi,
\label{selfconsistency}
\end{equation}
which is appropriate either if ${\mathcal A}(k^*)=0$ and the distribution is
symmetric in $\phi$, or if $\mathfrak{n}(k^*){\mathcal A}(k^*)$
is much larger than the (appropriately phased)
combined amplitudes of all the other modes.
The latter occurs if there is an instability that leads to
the formation of a pattern with wavenumbers of magnitude $k^*$.

Since $\rho(\phi ;0) =\rho(-\phi ;0)$, it follows that
${\mathcal A}(k^*)=0$ is always a solution.  To find other solutions
we expand the integral on the right side of Eq.~(\ref{selfconsistency})
around ${\mathcal A}(k^*) =0$,
\begin{equation}
\int_{-\infty}^\infty
\phi \rho \left( \phi ;{\mathcal A}(k^*) \right)
d\phi =b{\mathcal A}(k^*) +{\mathcal O}\left( {\mathcal A}^3(k^*)\right),
\end{equation}
where
\begin{equation}
b=\int_{-\infty}^\infty
\phi~\left. \frac{\partial \rho \left( \phi ;
{\mathcal A}(k^*) \right)}{\partial {\mathcal A}(k^*)}\right|_{{\mathcal
A}(k^*) =0} d\phi .
\end{equation}
It follows that self-consistent solutions different from
${\mathcal A}(k^*)=0$ are possible, and that
the loci that indicate
the appearance of these solutions satisfy $b=\mathfrak{n}(k^*)$, that is,
\begin{equation}
\frac{D_{1}}{\sigma ^{2}}\int_{-\infty}^\infty
\phi^2~\rho(\phi ;0)~
d\phi =1.
\label{loci0}
\end{equation}
The latter condition, which determines the phase diagram of the model,
can also be obtained by geometrical
arguments~\cite{vandenbroeck3,ojalvo}.
Equation~(\ref{loci0}) can be expressed in the
following algebraic form:
\begin{equation}
1=\frac{D_1{\mathbb K}_1\left( \frac{a+D_1}{4c\sigma^2}\right)
e^{\frac{a+D_1}{4c\sigma^2}}}{2\sigma^2c\sqrt{\pi }{\mathbb U}
\left(-\frac{1}{2},0,\frac{a+D_1}{2c\sigma^2}\right)},
\label{loci}
\end{equation}
where ${\mathbb U}(x,y,z)$ is the confluent hypergeometric
function and ${\mathbb K}_n(x)$ is the Bessel function of the second kind.

\begin{figure}
\includegraphics[width = 7cm]{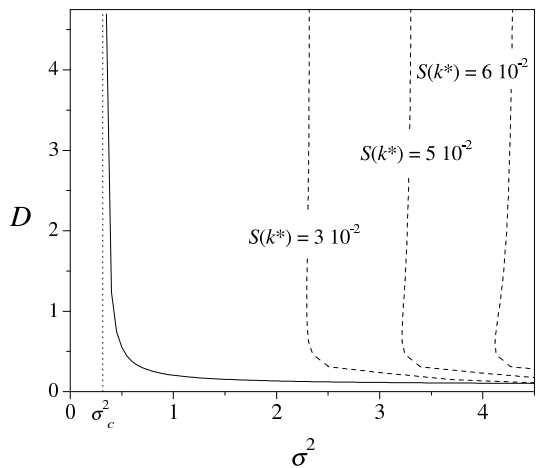}
\caption{Phase diagram of the model obtained from the modulated
mean field theory. The wide solid line indicates the transition
loci~(\ref{loci})
in $(D,\sigma^{2})$ space for $k_0=1$ ($k^*\approx 0.1035$),
$a=1$, and $c=3$.
Inside the ordered region where patterns develop we have
indicated some contour lines labeled by the value of the order
parameter $S(k^*)$ (see text). Note that as the coupling
goes to infinity, the
transition line tends to the value of the noise intensity where
the zero-dimensional system undergoes a noise-induced transition,
$\sigma_{c}^{2}=1/3$.}
\label{fig1}
\end{figure}

The derivation is not complete until we confirm, at least within
the approximations implemented in our mean field approach, that the
most unstable modes are indeed those of magnitude $k^*$, i.e., those
for which $\omega({\bm k}) =0$. This is easily ascertained as follows.
If we consider an ansatz state of the form (\ref{mean field}) but with a
different $k$, then we would arrive at a mean field equation
containing $\omega(k)$ explicitly:
\begin{equation}
\begin{aligned}
\dot{\phi}=&\Gamma (\phi) \left\{-(a-\omega(k))\phi
+D_{eff}(k) \left[\mathfrak{n}(k) {\mathcal A}(k) -\phi \right] \right\}
\\
&+\left[ \Gamma (\phi)\right]^{1/2}\xi(t),
\label{ec2}
\end{aligned}
\end{equation}
where $D_{eff}(k) =D_1+\omega(k)$ (cf. Eq.~(\ref{couplingapprox});
here the assumption has again been made that $\omega({\bm k})$ is well
approximated by $\omega(k)$.  Note that if $\omega(k)=0$ we recover
Eq.~(\ref{ec1}).
If we now follow the same steps leading to the condition (\ref{loci}) which
marks the boundary of pattern formation, we find that the {\em only}
modification is that the very first $D_1$ in the numerator is now replaced
with $D_{eff}(k)$.  The other occurrences of $D_1$ are not affected because
$D_1$ and $a$ are each shifted in such a way that their sum remains
unchanged.  Since $D_{eff}(k) < D_1$ for all $k\neq k^*$, it
follows that in order to satisfy Eq.(\ref{loci}) the noise intensity has to
be {\em greater} for other $k$.  In other words, while modes other than
those of magnitude $k^*$ may become unstable, they first do so at
higher values of the noise intensity. Conversely, for a given noise
intensity the coupling must be stronger to produce an instability of other
wavevectors.  Thus, as the noise intensity is
increased from zero for a given coupling, or the coupling is increased
from zero for a given noise intensity beyond $\sigma_c^2$, the first
and hence strongest instability occurs at $k^*$.

\begin{figure}
\includegraphics[width = 7cm]{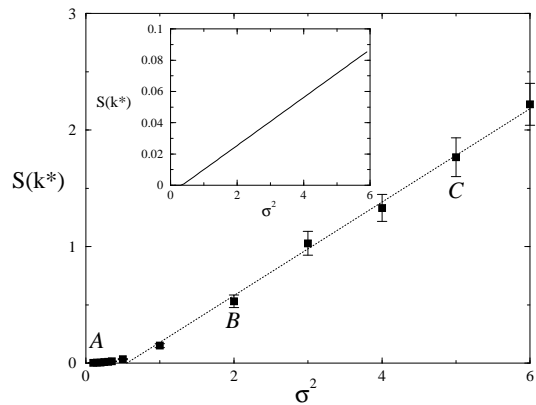}
\caption{
Order parameter as a function of the noise
intensity obtained from numerical simulations (squares)
for $a=1$, $c=3$, $D=5$, and $k_0=1$.  The error bars are indicated.
The dashed line indicates the linear interpolation of the numerical data used
to locate the transition point, $\sigma_{0}^{2} = 0.5\pm 0.1$.
The points labeled A ($\sigma^2=0.1$), B ($\sigma^2=2$), and
C ($\sigma^2=5$) correspond to the spatial structures shown
in Fig.~\ref{fig3}. Inset: mean field result, with transition
point at $\sigma_0^2=0.34955\pm 10^{-5}$.}
\label{fig2}
\end{figure}

\begin{figure}
\includegraphics[width = 4.75cm]{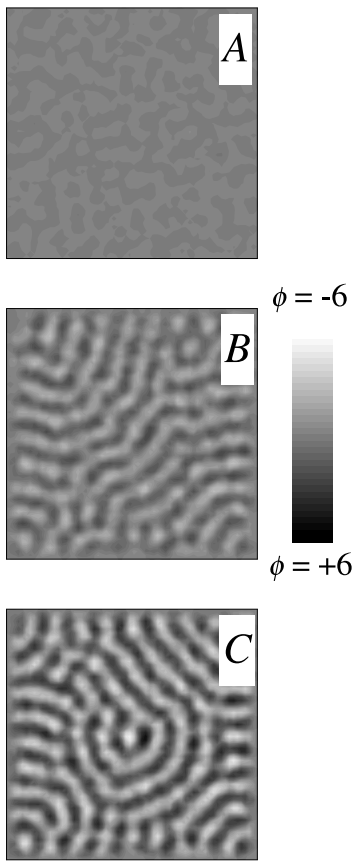}
\caption{
Density plots of the stationary field associated with the points
A (first panel), B (second), and C (third) of Fig.~\ref{fig2}.
Note that as predicted, for a constant value of the coupling, a pattern
develops as the intensity of the noise increases.
}
\label{fig3}
\end{figure}

Figure~\ref{fig1} shows the phase diagram of the model in the space
$(\sigma ^2,D) $ obtained from the numerical solution of
Eq.~(\ref{loci}) in two dimensions
with $\Delta x=1$, $a=1$, $c=3$, and $k_0=1$, which leads to
$k^*\approx 0.1035$.  Note that as $D$
goes to infinity the transition line moves to the value of the noise
intensity where the zero-dimensional stationary probability density becomes
bimodal, $\sigma _{c}^{2}=1/3$.

Since Eq.(\ref{model}) satisfies the inversion symmetry $\phi
\longleftrightarrow -\phi$, roll-shaped patterns are likely~\cite{cross}.
In order to characterize the structure that develops we use as an order
parameter the {\em total}
power spectrum at the most unstable modes,
\begin{equation}
S(k^*)= \sum_{\left\{\bm{k}^{\ast}\right\}}
\tilde{\phi}_{{\bm k}^*}
\tilde{\phi}_{-{\bm k}^*}
\end{equation}
where, as before,
$\tilde{\phi}_{\bm k}$ stands for the Fourier transform of the field
$\phi_{\bm r}$ and the sum is over all modes of magnitude $k^*$.
This parameter characterizes the transition from a laminar regime
(homogeneous)
to a convective regime (roll-like patterns) that we expect to obtain here.
Using our self-consistent solution then gives in the modified
mean field approximation the appropriately normalized relation
\begin{equation}
S(k^*)=\mathfrak{n}(k^*){\mathcal A}^2(k^*).
\end{equation}
In Fig.~\ref{fig1} we also present several contour lines indicating the
value of $S(k^*)$. Note that for a given value of the coupling $D$,
as noise
intensity increases, the order parameter $S(k^*)$ also increases,
that is, the stronger the noise, the larger is the amplitude of structures
associated with wavevector magnitude $k^*$.  Note also that for
sufficiently large noise intensity, for a given value of $\sigma^2$ there
is a non-monotonic behavior of $S(k^*)$ as a function of the coupling
strength, indicating that there exists a value of the coupling for which
the structures associated with $k^*$ exhibit maximum coherence.  At the same
time, the possible instability of other modes near $k^*$ may affect the
actual physical appearance of the system, so that these effects
may not be visually unequivocal.

Finally, we note that it is common in pattern formation discussions
to use the so-called {\em flux of convective heat}
\begin{equation}
J=\frac{1}{N^d}\sum_{\bm r} \phi^2_{\bm r} = \sum_{\bm k}
\tilde{\phi}_{{\bm k}} \tilde{\phi}_{-{\bm k}}
\label{flux}
\end{equation}
as an order parameter.  If our ansatz state were exact, then in the
patterned state the two are
identical, $J=S(k^*)$.  Differences point to the presence
of other unstable modes.

\begin{figure}
\includegraphics[width = 7cm]{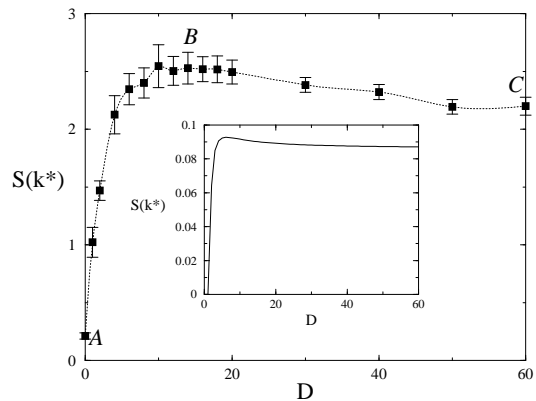}
\caption{
Order parameter as a function of the coupling strength
obtained from numerical simulations (squares)
for $a=1$, $c=3$, $\sigma^2=6$, and $k_0=1$.  The error bars are
indicated; the dashed line is a guide for the eye.
The points labeled A ($D=0$, uncoupled system), B ($D=14$), and
C ($D=60$) correspond to the spatial structures shown
in Fig.~\ref{fig5}. Inset: mean field result.}
\label{fig4}
\end{figure}

\begin{figure}
\includegraphics[width = 4.75cm]{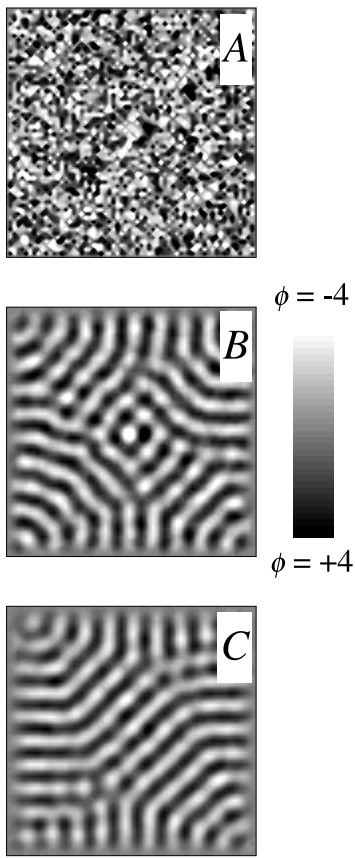}
\caption{
Density plots of the stationary field associated with the points
A (first panel), B (second), and C (third) of Fig.~\ref{fig4}.
Note that, as predicted, for a constant value of the noise, a pattern
develops as the coupling increases but then becomes less distinct as the
coupling increases further.
}
\label{fig5}
\end{figure}

\section{Numerical Simulations}
\label{numerical}

In order to check the analytical predictions of the modified
field theory, we perform
numerical simulations of Eq.~(\ref{model}) on a two dimensional square
lattice of $64\times 64$ cells and apply Dirichlet-Neumann
boundary conditions
commonly used in studies of fluids, namely, the field and the normal
component of the gradient are zero at the boundaries.
The relevant parameters have the following values in
our simulations: mesh size $\Delta x=1$ that gives the system length scale
$L=N\Delta x=64$, time step $\Delta t=0.001$, $k_0=1$, which leads to $k^*
\approx 1.035$ and an
{\em aspect ratio} $\Lambda =k^*L/2\pi \sim 10$.
The other parameters of the model are
again
taken as $a=1$ and $c=3$. Using these values, the onset
of bistability of the stationary one-site potential occurs at
$\sigma_{c}^{2}=a/c=1/3$. We expect the transition to pattern formation
to occur
for noise intensities near $\sigma _{c}^{2}$ if the coupling is
sufficiently strong.
According to the mean-field phase diagram Fig.~\ref{fig1}, $D=5$
is sufficiently large. We use this value of the coupling in subsequent
calculations and figures.  Although the initial conditions do not matter
for the final outcome, the outcome is reached more quickly if we start
from a target-like pattern of rings of width $k^*$.  We have ascertained
that other initial conditions, for example a random initial condition,
reach the stationary state albeit in a longer time.

We have numerically computed the order parameter for several noise
intensities, and the results are shown in
Fig.~\ref{fig2}. The order parameter is very small for low noise intensities
until the critical noise
intensity is reached, after which the order parameter grows linearly
with $\sigma^2$.  Our simulations lead to a critical value
$\sigma_0^2=0.5\pm 0.1$.  The mean field result, also shown in the figure,
predicts the ordering transition to occur at $\sigma_0^2=0.34955\pm
10^{-5}$ and also captures the linear transcritical behavior of
$S(k^*)$ with $\sigma ^{2}$, but predicts a slower growth of the
order parameter with noise intensity than the numerical simulations.
Nevertheless, the mean field theory clearly captures the full qualitative
behavior of the system, especially near the transition point.

We have also computed the flux according to Eq.~(\ref{flux}) for
all the cases for which we have presented the order parameter.
We find that both predict essentially the same critical noise value for
the appearance of patterns, and that for our simulation parameters
the flux is fairly consistently
twice as large as the order parameter (because of the contributions of
other modes to  the former).  This consistency would indicate that
even well beyond the transition point the most unstable modes dominate the
flux.

The spatial pattern that emerges as $\sigma ^{2}$ increases is
illustrated in
Fig.~\ref{fig3} using density plots of the stationary value of the field
for the noise intensities labeled A, B, and C in Fig.~\ref{fig2}: $\sigma
_{A}^{2}=0.1$ (weak noise, no pattern), $\sigma _{B}^{2}=2$
(rolls are visible), and $\sigma _{C}^{2}=5$ (strong noise,
distinctive pattern).

In order to ascertain the non-monotonic behavior of the order parameter
with coupling strength at a given noise intensity,
in Fig.~\ref{fig4} we present the results for a fixed noise
intensity $\sigma^2=6$.  Simulations are indicated by squares, and
the inset shows the results predicted by the theory. While there are again
quantitative differences, the qualitative agreement is evident: both
curves show a clear non-monotonic behavior.
The density plots of the field associated with points $A$, $B$,
and $C$ in Fig.~\ref{fig5} confirm this behavior. Of course even as the
amplitudes of wavevectors of magnitude $k^*$ are decreasing, the
amplitudes of other nearby wavevectors may be growing.  Therefore, while
the lightening of pattern $C$ on the gray scale reflects the decrease in
the order parameter, a visual perception of some loss of distinctness
could be due to an admixture of other wavevectors.

\section{Conclusions}
\label{conclusions}

We have shown by means of a modulated mean field
approximation and numerical
simulations that the mechanism for
noise-induced phase transitions introduced by Iba\~nes
{\em et al.}~\cite{ibanes}
can be extended to pattern formation
phenomena. In contrast with previous work on noise-induced patterns,
no short-time instability is required to generate these spatial structures.
As a consequence, this transition is independent of the noise
interpretation, as has been shown  for noise--induced phase
transitions~\cite{oliver2}.
For example, for the It$\hat{\rm o}$ interpretation we have checked numerically
that this model exhibits
the same noise-induced patterns, but at a lower critical noise intensity.
Furthermore, no re-entrance phenomenon occurs as the fluctuations
grow in intensity. Indeed, in our system stronger noise leads to
increasingly ordered structures. The mean field theory has allowed us to
characterize the model by computing its phase diagram. We have also
performed numerical simulations that confirm the qualitative
validity of the theoretical analysis.  We find that
as the intensity of the fluctuations increases, a rotationally symmetric
roll-shaped pattern appears.  The pattern is characterized by the most
unstable modes of the system, those with wavevector of a magnitude
$k=k^*$ that is explicitly predicted by the mean field analysis.
Both the theoretical and numerical analyses show that as the
coupling between sites goes to infinity, the transition
to pattern formation
occurs at the same point where the zero-dimensional system presents a
noise-induced transition. We have also shown that for sufficiently
strong noise intensity the order parameter for the system is non-monotonic
as a function of the coupling strength.
Thus, for sufficiently strong noise there exists an optimal
value of the coupling such that the patterns of characteristic size
$2\pi/k^*$ are maximally coherent.

\begin{acknowledgments}
This work was partially supported by the National Science Foundation under
grant No. PHY-9970699,
by MECD-Spain Grant No. EX2001-02880680, and by MCYT-Spain Grants
Nos. BFM2001-0291 and BFM2000-0624.
\end{acknowledgments}

\appendix

\section{Implementation of the Modified Mean Field Theory}
\label{a}

We begin by exhibiting in some detail the dependences associated with the
ansatz field~(\ref{mean field}) and the action of ${\mathcal L}$ on it.
For example, for an ${\bm r}'$ that
is $m$ lattice sites away from ${\bm r}=(x_1, x_2, \ldots, x_d)$
in the direction $j$ the ansatz reads
\begin{equation}
\phi(x_1,x_2,\ldots,x_j+m\Delta x,\ldots,x_d)
=\sum_{\left\{\bm{k}^{\ast}\right\}}\mathcal{A}(k^*)
\cos(m\Delta x k_j).
\label{axis}
\end{equation}
For an  ${\bm r}'$ that is in the immediate positive diagonal location
away from ${\bm r}$ we have
\begin{equation}
\begin{aligned}
\phi(x_1+\Delta x,& x_2+\Delta x,\ldots,x_j+\Delta x,\ldots,x_d +\Delta x)
\\
&=\sum_{\left\{\bm{k}^{\ast}\right\}}\mathcal{A}(k^*)
\cos[\Delta x (k_1 +k_2+\cdots+k_d)].
\label{diagonal}
\end{aligned}
\end{equation}
Next, to apply the discrete version (\ref{l-discrete}) of ${\mathcal L}$
we must elucidate
the effect of the operators
$\left[ \sum_{i=1}^{d}\sinh^2\left(\frac{\Delta x}{2}\frac{\partial}
{\partial x_i}\right)\right]^n$ on the field
$\phi_{\bm{r}}$ for $n=1,2$. With $n=1$, we use the relation $2\sinh^2(y/2)=
[\cosh(y)-1]$ and note that
\begin{equation}
\begin{aligned}
&\sum_{i=1}^d\cosh\left(\Delta x\frac{\partial}{\partial x_i}\right)
\phi_{\bm{r}} =\frac{1}{2}\left[\phi(x_1+\Delta
x,x_2,\ldots,x_j,\ldots,x_d) \right. \\
&~~~~~~~ +\phi(x_1-\Delta x,x_2,\ldots,x_j,\ldots,x_d) \\
&~~~ \ldots+\phi(x_1,x_2,\ldots,x_j+\Delta x,\ldots,x_d) \\
&~~~~~~~+\phi(x_1,x_2,\ldots,x_j-\Delta x,\ldots,x_d)  \\
&~~~ \ldots+\phi(x_1,x_2,\ldots,x_j,\ldots,x_d+\Delta x)\\
&\left. ~~~~~~~+\phi(x_1,x_2,\ldots,x_j,\ldots,x_d-\Delta x)
\right] .
\end{aligned}
\end{equation}
By using Eq.~(\ref{axis}) in this last equation we obtain
\begin{equation}
\sum_{i=1}^d\cosh\left(\Delta x\frac{\partial}{\partial x_i}\right)
\phi_{\bm{r}}=\sum_{\left\{ \bm{k}^{\ast}\right\}}\mathcal{A}(
k^*)\sum_{i=1}^d\cos(k_i\Delta x) .
\label{cosh}
\end{equation}
As for $n=2$, we note that
$4\sinh^2(y/2)\sinh^2(z/2)=[\cosh(y)-1][\cosh(z)-1]$ and, in turn,
$\cosh(y)\cosh(z)=\frac{1}{2}[\cosh(y+z)+\cosh(y-z)]$.  The latter
combination leads to contributions that involve
both forward and backward
translations in different spatial directions. This is easily
visualized by noting explicitly that
\begin{equation}
\begin{aligned}
&\left[\sum_{i=1}^d  \cosh \left(\Delta x\frac{\partial}{\partial x_i}
\right) \right] ^{2} 
\\
&~~~~~~~=\frac{1}{2}\left[\sum_{i,j=1}^{d}\cosh\left(
\Delta x\left(\frac{\partial}{\partial x_i}+\frac{\partial}{\partial x_j}
\right) \right) \right. \\
&~~~~~~~~~ + \left. \cosh\left( \Delta x\left(\frac{\partial}{\partial
x_i}-\frac{\partial}{\partial x_j}\right) \right) \right] .
\end{aligned}
\end{equation}
Notice that for the $d$ cases where with $i=j$, the second term on the right
hand side leaves the field at the original site $\bm{r}$. The field at the
original site is not represented by the ansatz assumption, and therefore
we must subtract the
$d$ ``spurious" terms produced by the ansatz state
and add $d$ times the field $\phi_{\bm{r}}$. This procedure leads to,
\begin{equation}
\begin{aligned}
&\left[ \sum_{i=1}^d\cosh\left(\Delta x\frac{\partial}{\partial x_i}
\right) \right]^{2}\phi_{\bm{r}}
\\
&~~~=\frac{d}{2}\phi_{\bm{r}}
+\sum_{\left\{\bm{k}^{\ast}\right\} }\mathcal{A}(k^*)
\left[ \left(\sum_{i=1}^d\cos(k_i\Delta x) \right)^{2}-\frac{d}{2}\right] .
\label{cosh2}
\end{aligned}
\end{equation}
Note that we have taken advantage of the directional insensitivity of
$k^*$.

Use of Eqs.~(\ref{cosh}) and (\ref{cosh2}) in
Eq.~(\ref{l-discrete})
then leads to the following approximation for the term containing the
SH coupling operator:
\begin{equation}
\begin{aligned}
\mathcal{L}\phi_{\bm{r}}=&D_1\left( \sum_{\left\{\bm{k}^{\ast
}\right\} }\mathcal{A}(k^*) -\phi_{\bm{r}}\right)
+\sum_{\left\{ \bm{k}^{\ast}\right\} }\mathcal{A}(k^*)
\omega(k^*)
\\
=& D_1\left( \sum_{\left\{\bm{k}^{\ast
}\right\} }\mathcal{A}(k^*) -\phi_{\bm{r}}\right) ,
\label{couplingapprox}
\end{aligned}
\end{equation}
where the last term vanishes because $\omega(k^*)=0$, and
where $D_1=D\left[\left(\frac{2d}{\Delta x^2}-k_0^2\right)
^2+\frac{2d}{\Delta x^4}\right]$.

Finally, since the summand in Eq.~(\ref{couplingapprox}) is
independent of the direction
of the ${\bm k}^*$, the sums simply give the number of terms in the sum
(or the appropriate integral form) times the summand.  Simulations always
involve a {\em finite} system of $N^d$ sites (i.e., of
volume $(N\Delta x)^d$), so that the allowed
modes themselves form a discrete set, with each component separated from
the next one by an interval $\Delta k = \frac{2\pi}{N \Delta x}$.
One way to count the number of modes $\mathfrak{n}(k^*)$
in the sum is to construct a ring
of radius $\min k^*$ (which we shall simply call $k^*$ following the
discussion surrounding Eqs.~(\ref{mink}) and (\ref{maxk})) of
thickness $\Delta k\equiv 2\pi/N\Delta x$, and to consider all the modes that
lie in this ring.  We can then estimate this number by calculating
the number of cells of volume $(2\pi/N \Delta x)^d$ in the ring:
\begin{equation}
\mathfrak{n}(k^{\ast}) =\frac{d\pi^{d/2}}{\Gamma(d/2+1)}
\left(\frac{Nk^*}{2\pi}\right)^{d-1}.
\end{equation}
Although variations in the particular way of counting are
possible, for sufficiently large $N$ the differences are small.
Thus we finally arrive at the mean field approximation
\begin{equation}
\mathcal{L}\phi_{\bm{r}}=D_1\left[\mathfrak{n}(k^{\ast})
\mathcal{A}(k^{\ast})-\phi_{\bm{r}}\right].
\end{equation}

\end{document}